\documentclass[english, 11pt, a4paper]{article}
\usepackage[utf8]{inputenc}
\usepackage{babel}
\usepackage{graphics}
\usepackage{graphicx}
\usepackage{texdraw}
\usepackage{epsfig}
\usepackage{float}
\usepackage{amsmath}
\usepackage{amssymb}
\usepackage{enumerate}
\usepackage[all]{xy}
\usepackage{fullpage}
\usepackage[small]{caption}
\begin{document}
%LaTeX Commands%

\def\lsim{\mathrel{\rlap{\lower4pt\hbox{\hskip1pt$\sim$}}
    \raise1pt\hbox{$<$}}}                % less than or approx. symbol

\def\gsim{\mathrel{\rlap{\lower4pt\hbox{\hskip1pt$\sim$}}
    \raise1pt\hbox{$>$}}}                % greater than or approx. symbol

\newcommand\cotanh{ {\, \rm cotanh \, } }
\newcommand{\dslash}[1]{#1\kern-0.70em/}
\newcommand{\minislash}[1]{#1\kern-0.50em/}

\title{Progress in numerical simulations of systems with a
  $\theta-$vacuum like term: The two and three-dimensional Ising model
  within an imaginary magnetic field}

\author{Vicente Azcoiti, Eduardo Follana and Alejandro Vaquero,\\
\\
	{\small \emph{Departamento de Física Teórica, Facultad de Ciencias, Universidad de Zaragoza}}\\{\small \emph{C/ Pedro Cerbuna 12, E-50009, Zaragoza (Spain)}}}
\date{\today}
\maketitle
\setcounter{page}{1}

\abstract{Using the approach developed in \cite{REFVIC2}, we succeeded
  in reconstructing the behaviour of the antiferromagnetic Ising model
  with imaginary magnetic field $i\theta$ for two and three dimensions
  in the low temperature regime. A mean-field calculation, expected to
  work well for high dimensions, is also carried out, and the
  mean-field results coincide qualitatively with those of the two- and
  three-dimensional Ising model. The mean field analysis reveals also
  a phase structure more complex than the one expected for $QCD$ with
  a topological $\theta-$term.}

\section{Introduction}

Quantum field theories with complex actions are systems as interesting
as difficult to analyze: on one hand, the complex action usually gives
rise to a severe sign problem, which prevents computer simulations to
be performed; on the other hand, the number of known, exactly solvable
models with complex actions is quite limited. And yet, there are very
important models in this class, for instance, QCD with a $\theta-$term
or at finite baryon density, whose solution might lead to an
explanation of the strong $CP$ problem and to the understanding of the
rich phase structure expected for matter at high pressures. In
condensed matter physics Haldane showed \cite{haldane} that chains of
quantum half-integer spins with antiferromagnetic interactions are
related to the two-dimensional $O(3)$ nonlinear sigma model with a
topological term at $\theta=\pi$, and conjectured that such model
presents a second order phase transition at $\theta=\pi$, keeping its
ground state $CP$ symmetric. These are some of the reasons why a great
effort is being invested in developing new algorithms, capable of
dealing with complex actions.

For the particular case of $\theta-$vacuum systems, the partition
function in the presence of a $\theta$ term is periodic and,
conveniently normalized, can be decomposed in sectors of different
topological charge $n$ (or density of topological charge $x_n=n/V$),
as
\begin{equation}
Z_V\left(\theta\right) = \sum_n p_V\left(n\right)e^{i\theta n} =
\sum_n e^{-Vf_V\left(x_n\right)}e^{i\theta Vx_n},
\label{Is-19b}
\end{equation}
which resembles the Fourier transform of the probability distribution
function (p.d.f.) of the topological charge at $\theta=0$. The
probability of the topological sector $n$ is therefore given by
$p_V\left(n\right)$, and this quantity can, a priori, be measured from
numerical simulations. Unfortunately, this is a very difficult task,
for several reasons:
\begin{enumerate}[i.]
\item The precision in a numerical simulation is limited by
  statistical fluctuations. Thus the measurement of
  $p_V\left(n\right)$ suffers from errors.
\item Small errors in $p_V\left(n\right)$ induce huge errors in the
  determination of $Z_V\left(\theta\right)$, as this quantity is
  exponentially small, $Z_V\approx e^{-V}$, due to the sign problem.
\item Even if we were able to evaluate $p_V\left(n\right)$ with
  infinite accuracy, the terms on the sum \eqref{Is-19b} differs by
  many orders of magnitude (from 1 to $e^{-V}$).
\end{enumerate}

In fact, the different groups that have tried to determine with high
precision the p.d.f. of the topological charge either by standard
simulations, or by more sophisticated methods (re-weighting or
multibinning techniques), have found artificial phase transitions in
the $U(1)$ and $CP^N$ models. The reason behind these ghost
transitions is the flattening of the free energy for $\theta-$values
larger than a certain threshold. In \cite{Ref3,Ref7}, this threshold
is roughly evaluated, and the flattening behaviour explained. The
conclusion is clear: a reliable computation of the order parameter (or
the free energy) for all values of $\theta$ from the direct
measurement of the p.d.f. of the topological charge is not feasible
due to the huge statistics required.

That is why other approaches \cite{wolff, wiese, verba, luigi,
  massimo, pepe, japon, alessandro} (or at least, serious refinements
of the standard approach) should be considered. In \cite{PdfIsing}, a
remarkable breakthrough was achieved, pushing the threshold of
$\theta$ to its limit $\theta=\pi$. The method is based on the observation that,
since all the coefficients entering in the right
hand side of equation \eqref{Is-19b} are positive at purely imaginary
 values of $\theta$, the free energy is given in the thermodynamic limit by the
saddle point
\begin{equation}
f'\left(x\right) = h,
\label{Is-19g}
\end{equation}
where $f\left(x\right)$ represents $f_V\left(x_n\right)$ as
$V\rightarrow\infty$, and $h$ stand for a purely imaginary $\theta$
 field $\theta = -ih$.

Then, the function was fitted to a ratio of polynomials, and
integrated analytically to obtain $f(x)$. This step is essentially
different to what other groups proposed, and it solved the problem of
the $\theta$ threshold in some systems. Finally, a multi-precision
algorithm is used to calculate the partition function directly from
\eqref{Is-19b}, using the function $f(x)$.

It was demonstrated numerically in \cite{PdfIsing} that the errors in
the reconstructed $f\left(x\right)$ using this method were highly
correlated. This was a great advantage, for the errors, propagated to
the exact free energy density, became almost constant, and these
errors amounted to an irrelevant constant in the free energy. These
ideas were successfully tested \cite{PdfIsing} in the one-dimensional
Ising model and the two-dimensional $U(1)$ model, and the method was
used to predict the behaviour of the $CP^3$ model. Furthermore, in
reference \cite{SummaryIsing} the continuum $\theta$ dependence of
$CP^9$ --a confining and asymptotically free quantum field theory--
was fully reconstructed. Data collapsed for different couplings within
percent level and this evidence for scaling at non-zero $\theta$ is
the strongest indication that the $CP$ symmetry is spontaneously
broken in the continuum, as predicted by the large $N$ expansion.

The results were impressive by that time, solving completely the
problem of the flattening of the order parameter beyond the critical
value of $\theta$. The key of this success was the aforementioned
correlation among the errors: tests performed using the same method
and adding an apparently negligible $0.1\%$ uncorrelated random error
to the measured free energy $f\left(x\right)$ led to
disaster. Nonetheless, if the error was correlated, it could be as
large as $50\%$, and the final result would be quite reasonable.

However, this method is not yet generally applicable. The flattening
can appear --and in fact, does-- whenever the behaviour of the order
parameter is not monotonous. This seems to be a general rule. The
flattening was first observed in a simple toy model which featured
symmetry restoration at $\theta=\pi$,
\begin{equation}
f\left(\theta\right) = \ln\left(1 + A\cos\theta\right).
\label{Is-19j}
\end{equation}
For this model, the order parameter vanishes only at $\theta = 0,\pi$
\begin{equation}
im\left(\theta\right) = \frac{A\sin\theta}{1 + A\cos\theta},
\label{Is-19k}
\end{equation}
but the method predicted an almost flat behaviour, slightly
increasing, beyond the point $\theta=\frac{\pi}{2}$.

On the whole, although the method proposed in \cite{PdfIsing}
represented a large improvement over what existed at that point, it
was clear that another approach was necessary, and that is how the
method described in the following section was created \cite{REFVIC2}.

\section{One-dimensional Ising model}

The Ising model is the simplest model describing ferromagnets, but it
is also a good theoretical laboratory to test new algorithms. It is
easy to code efficiently on the computer, and simulations are very
fast, allowing the generation of big statistics, even on large
lattices. The one-dimensional model in a magnetic field is exactly
soluble, which allows us to check our results against the exact
solution. We can also identify in some sense magnetization and
topological charge in this model, and regard an imaginary external
magnetic field as a $\theta$ term in the action. Finally, from the
numerical point of view, it is even more challenging\footnote{As the
  following sections show, the phase diagram of the Ising model within
  an imaginary magnetic field is richer than the one expected for QCD
  in presence of a $\theta-$vacuum term.}  than other complex systems
suffering from the sign problem, such as lattice QCD with a
$\theta-$term, yet it remains more accessible. Therefore, it is a good
idea to check the goodness of any algorithm in this toy model, prior
to its application to more physically interesting systems.

The hamiltonian of the one-dimensional Ising model with nearest
neighbours coupling $J$ and external magnetic field $B$ is
\begin{equation}
H\left(\{s_i\},J,B\right)=-J \sum^N_i s_i s_{i+1} - B \sum^N_i s_i.
\label{Is-10}
\end{equation}
Defining reduced couplings $F = J / (KT), h = 2 B / (KT)$, the density
of free energy is given by
\begin{equation}
f\left(F,h\right) = F + \ln\left(\cosh\frac{h}{2} + \sqrt{e^{-4F} +
  \sinh^2\frac{h}{2}}\right),
\label{Is-17}
\end{equation}
where $f\left(F,h\right) = 1/V\ln Z$ represents the free energy. It 
is quite remarkable that the Ising model within an imaginary
external field (i.e., for $h=-i\theta$ with $\theta\in\mathbb{R}$) is 
not properly defined with ferromagnetic couplings \cite{China}. 
Setting $F>0$ we find that the free energy \eqref{Is-17} becomes 
undefined for certain values of $\theta$, for the argument of the 
logarithm may vanish if $F>0$. Hence, we will deal with the 
antiferromagnetic Ising model ($F<0$) from now on.

In systems with an even number of spins, the quantity
$\frac{M}{2}=\frac{1}{2}\sum^N_{j=1} s_j$ is an integer taking any value between
$-N/2$ and $N/2$. From equation \eqref{Is-17}, the mean density of
magnetization is
\begin{equation}
\langle m\rangle = \frac{\partial f}{\partial\frac{h}{2}} =
\frac{\sinh\frac{h}{2}}{\sqrt{e^{-4F} + \sinh^2\frac{h}{2}}}.
\label{Is-18}
\end{equation}
Equation \eqref{Is-18} for the magnetization is completely general, in
particular it is valid for the case of a pure imaginary magnetic field
$h = i\theta$. For $\theta = \pi$, the $Z_2$ symmetry is restored (the
magnetic field amounts to a sign $\sigma$, depending on the
`topological charge' $M/2$ of the configuration $\sigma=e^{i\pi M/2}$;
this sign is invariant under a $Z_2$ transformation). Then, the
question is whether the $Z_2$ symmetry is spontaneously broken or
not. Substituting $h\rightarrow i\theta$ in \eqref{Is-18}, we get
\begin{equation}
\langle m\rangle = \frac{i\sin\frac{\theta}{2}}{\sqrt{e^{-4F} - \sin^2
    \frac{\theta}{2}}}.
\label{Is-19}
\end{equation}
\begin{figure}[h!]
\centerline{\input{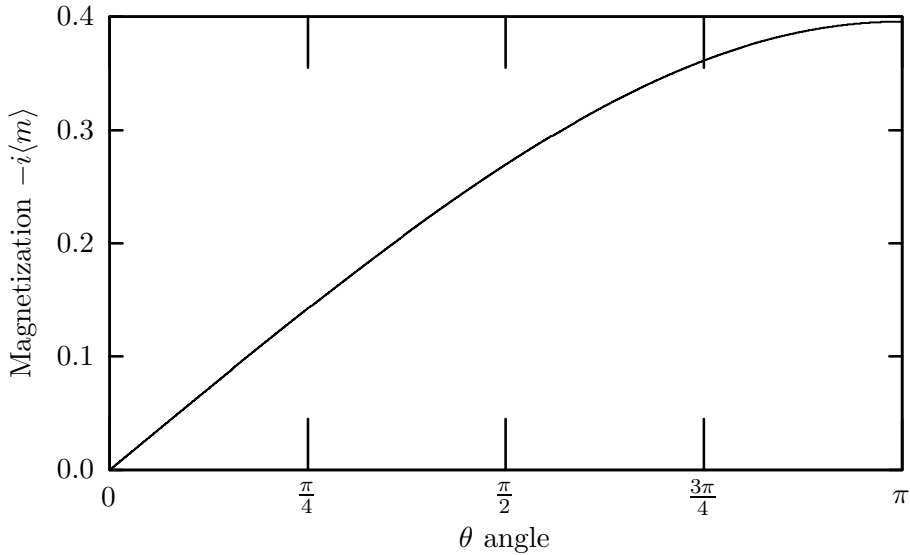}}
\caption{\label{iMag-I}Magnetization density as a function of the
  $\theta$ angle for the one-dimensional Ising model. $F$ was set to
  $F=-0.50$.}
\end{figure}
Thus the magnetization takes a non-zero expectation value for the
one-dimensional Ising model at $\theta = \pi$, a fact that indicates
spontaneous symmetry breaking (see fig. \ref{iMag-I}).

The Ising model in one dimension has a striking scaling property. Let
us define the variable $z = \cosh\frac{h}{2}$, and compute the ratio
$\frac{\langle m\rangle}{\tanh\frac{h}{2}}$:
\begin{equation}
y\left(z\right) = \frac{\langle m\rangle}{\tanh\frac{h}{2}} =
\frac{\left(e^{-4F} - 1\right)^{-\frac{1}{2}} z}{\sqrt{\left(e^{-4F} - 1\right)^{-1}z^2 + 1}}
\label{Is-20}
\end{equation}
Therefore $y(z)$ depends on $z$ and $F$ only through the
combination $\left(e^{-4F} - 1\right)^{-\frac{1}{2}}z$. Due to this simplified
dependency on $F$ and $h$, the transformation
\begin{equation}
y_\lambda\left(z\right) = y\left(e^{\frac{\lambda}{2}}z\right)
\label{Is-21}
\end{equation}
with $\lambda\in\mathbb{R}$ is equivalent to a change in the reduced coupling $F$, or in the
temperature of the model. The interesting point here is the fact that
for negative values of $\lambda$, this transformation can take the variable $z =
\cosh \frac{h}{2}$ to the range $0 < z < 1$, therefore $z = \cos
\frac{\theta}{2}$, corresponding to an imaginary field. This means
that we can measure $y\left(z\right)$ for imaginary values of the
magnetic field by mean of numerical simulations at real values of $h$,
which are free from the sign problem.

In order to check if a similar scaling property still holds for other
systems, let us assume that $y\left(z,F\right) =
y\left(g\left(F\right)z\right)$, then
\begin{equation}
\frac{\frac{\partial y}{\partial F}}{\frac{\partial y}{\partial z}} =
\frac{g'\left(F\right)z}{g\left(F\right)}
\label{Is-23}
\end{equation}
To have scaling, the ratio $\frac{\partial y}{\partial
  F}/z\frac{\partial y}{\partial z}$ should be independent of $h$.

\begin{figure}[h!]
\centerline{\input{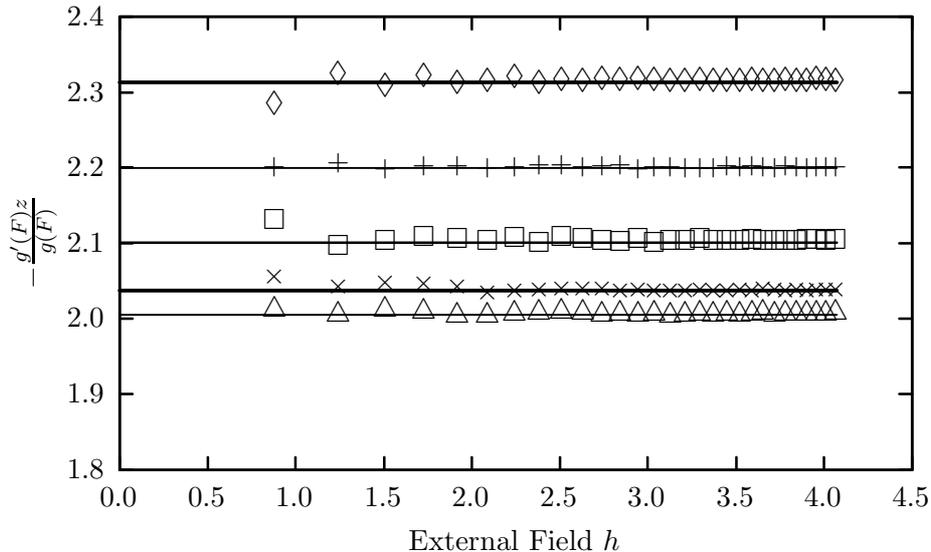}}
\caption{\label{Mi1d}Ising's scaling check along formula
  \eqref{Is-23}. The continuous lines represent the analytical result,
  while the crosses stand for the numerical data. We performed short
  simulations (only $\sim100000$ iterations) for several values of $F$
  in a $L=100$ lattice. Errors are smaller than symbols.}
\end{figure}

For the one-dimensional Ising model the simulations exhibit a constant
ratio over a large range of fields $h$ (see Fig. \ref{Mi1d}).

\begin{figure}[h!]
\centerline{\input{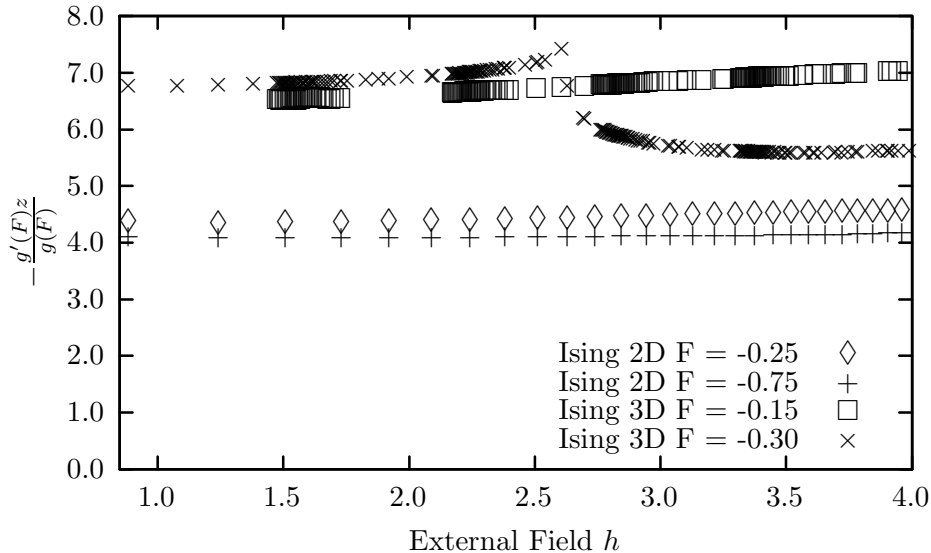}}
\caption{\label{Mind}Ising's results in 2D and 3D. The scaling is
  approximate in 2D and 3D for low values of the field. The statistics
  of the simulations are $100000$ iterations, and the lattice lengths
  are, for 2D $L=50$ and for 3D $L=25$. Errors are smaller than
  symbols.}
\end{figure}

Unfortunately, this property is exclusive of the one-dimensional
case. For two dimensions the ratio shows a slightly dependence on the
reduced magnetic field. For three dimensions, the dependence becomes a
bit more pronounced. The peak in Fig. \ref{Mind} is produced by the
antiferromagnetic-ferromagnetic phase transition\footnote{The 
antiferromagnetic Ising model displays, for strong enough couplings, 
a phase transition at non-zero external magnetic field : the 
spin-coupling tries to put the system in an antiferromagnetic state, 
whereas the external field tries to order the spins in a 
ferromagnetic fashion. As the value of the external field increases, 
the ferromagnetic behaviour takes over.}.

In any case, the dependence on the external field for these two models is mild for small values of the field $h$, far from the transition point, and large values of $|F|$ (for low temperatures). We will see later that this property becomes very relevant when dealing with asymptotically free theories.

\section{\label{MiIs}Computing the order parameter for 
imaginary magnetic fields}

Although the exact scaling property is absent in higher dimensions, we
can still take advantage from the methodology it suggests. For the
one-dimensional case, a measurement of the order parameter produced at
the point $\left(F,z\right)$ is equivalent to a measurement done at
$\left(F',z'\right)$ if the following relationship holds
\begin{equation}
g\left(F\right)z = g\left(F'\right)z',$$
$$g\left(F\right) = \left(e^{-4F} - 1\right)^\frac{1}{2}.
\label{Is-28}
\end{equation}
This way, and choosing carefully the value of $F$, a simulation
performed at a real value of the reduced magnetic field $z\geq 1$ is
equivalent to another simulation performed at imaginary values of $h$
(where $z < 1$).

The procedure to find out the order parameter at imaginary values of
the reduced magnetic field relies on scaling transformations
\cite{REFVIC2}. We define the function $y_\lambda\left(z\right)$ as
\begin{equation}
y_\lambda\left(z\right) = y\left(e^{\frac{\lambda}{2}}z\right).
\label{Is-29}
\end{equation}
For negative values of $\lambda$, the function
$y_\lambda\left(z\right)$ allows us to calculate the order parameter
$\left(\tanh\frac{h}{2}\,y\left(z\right)\right)$ below the threshold
$z = 1$. If $y\left(z\right)$ is non-vanishing for any positive
$z$\footnote{Even though the possibility of a vanishing
  $y\left(z\right)$ for some value $z>0$ can not be excluded
  completely, it does not happen for any of the analitically solvable
  models we know.}, then we can plot $y_\lambda/y$ against
$y$. Furthermore, in the case that $y_\lambda/y$ is a smooth function
of $y$ close to the origin, then we can rely on a simple extrapolation
to $y=0$. Of course, a smooth behaviour of $y_\lambda/y$ can not be
taken for granted; however no violations of this rule have been found
in the exactly solvable models.

The behaviour of the model at $\theta = \pi$ can be ascertained from
this extrapolation. At this $\theta$ the model has the same $Z_2$
symmetry as in the absence of field. We define a critical exponent
$\gamma_\lambda$
\begin{equation}
\gamma_\lambda = \frac{2}{\lambda}\ln\left(\frac{y_\lambda}{y}\right)
\label{Is-30}
\end{equation}
As $z\rightarrow0$, the order parameter
$\tan\frac{\theta}{2}\,y\left(\cos\frac{\theta}{2}\right)$ behaves as
$\left(\pi-\theta\right)^{\gamma_\lambda-1}$. Therefore, a value of
$\gamma_\lambda = 1$ implies spontaneous symmetry breaking at $\theta
= \pi$. A value between $1 < \gamma_\lambda < 2$ signals a second
order phase transition, and the corresponding susceptibility
diverges. Finally, if $\gamma_\lambda = 2$, the symmetry is realized
(at least for the selected order parameter), there is no phase
transition and the free energy is analytic at
$\theta=\pi$.\footnote{Other possibilities are allowed, for instance,
  any $\gamma_\lambda > 1,\quad\gamma_\lambda\in\mathbb{N}$ leads to
  symmetry realization for the order parameter at $\theta=\pi$ and to
  an analytic free energy. If $\gamma_\lambda$ lies between two
  natural numbers, $p < \gamma_\lambda < q,\quad p,q\in\mathbb{N}$,
  then a transition of order $q$ takes place.}

\begin{figure}[h!]
%\centerline{\includegraphics[width=0.85\textwidth]{Pictures/ThetaM2.2.eps}}
  \centerline{\includegraphics[angle=90,width=0.7\textwidth]
    {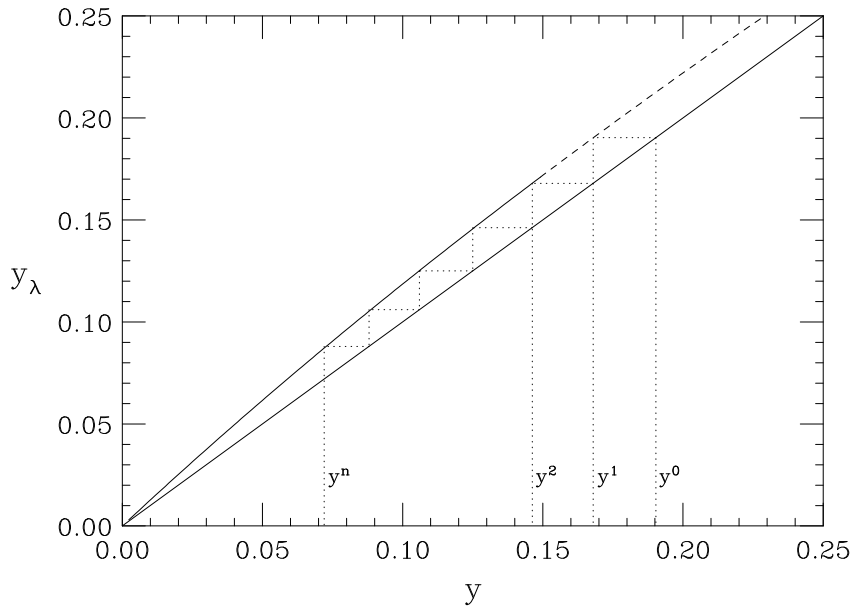}}
\caption{\label{It-M2}Iterative method used to compute the different
  values of $y(z)$. $y_\lambda$ is ploted as a function of $y$ using a
  dashed line in the region where direct measurements are available,
  and a continuous line in the extrapolated region. The straight
  continuous line represents $y_\lambda=y$.}
\end{figure}

We can take the information contained in the quotient
$\frac{y_\lambda}{y}\left(y\right)$, and calculate the order parameter
for any value of the imaginary reduced magnetic field $h=-i\theta$
through an iterative procedure \cite{REFVIC2}. The outline of the
procedure is the following:
\begin{enumerate}[i.]
\item Beginning from a point $y\left(z_i\right) = y_i$, we find the
  value $y_{i+1}$ such that $y_\lambda=y_i$. By definition,
  $y_{i+1}=y\left(e^{\frac{-\lambda}{2}}z_i\right)$.
\item Replace $y_i$ by $y_{i+1}$, to obtain
  $y_{i+2}=y\left(e^{-\lambda}z_i\right)$.
\end{enumerate}
The procedure is repeated until enough values of $y$ are know for
$z<1$ (see Fig. \ref{It-M2}). This method can be used for any model, 
as long as our assumptions of smoothness and absence of singular
behaviour are verified during the numerical computations. Indeed the
reliability of our approach in practical aplications is better when the
following two points are well realized:
\begin{enumerate}[a.]
\item $y(z)$ takes small values for values of $z$ of order 1.
\item The dependence on $y$ of the functions $y_\lambda/y$ and $\gamma_\lambda$
is soft enough to allow a reliable extrapolation.
\end{enumerate}
In the one-dimensional model these two properties are realized in the low
temperature regime (see equation \ref{Is-20} and Fig. 5),
but the two and three-dimensional models, notwithstanding they do not verify a
perfect scaling law as in the
one-dimensional case, they also show a very good behaviour (see
Figs. 7, 8). Indeed the relevant feature, at least in what concerns
point $a$, is that, at low temperatures, the magnetic susceptibility at small
values of the real external magnetic field $h$, which is essentially $y(z)$,
takes small values; and this is also true for any dimension. In the more
interesting case of asymptotically free models, the analogue of the magnetic
susceptibility is the topological susceptibility, and it is well known that
topological structures are very suppresed near the continuum limit. Therefore,
and on qualitative grounds, we expect a much better implementation of our
method in the low temperature regime of the Ising model and near the continuum
limit of asymptotically free theories. A check of this statement for the Ising
model is the content of this article, and concerning asymptotically free
models, the method was successfuly applied to the analysis of the continuum
$\theta-$dependence of $CP^9$ \cite{SummaryIsing}, showing a very good
realization of points $a$ and $b$. In the more general cases we should find out
whether the model complies with these two points or not, and pleasant
surprises are not
excluded. Indeed, we checked in \cite{HaldaneMl} that the conditions for
the aplication of our approach to $CP^1$ also hold,
 and this allowed us to verify the
Haldane conjecture and the relevant universality class of the
non-linear $O(3)$ $\sigma$-model in two dimensions.

\section{Numerical results}

\begin{figure}[h!]
\centerline{\input{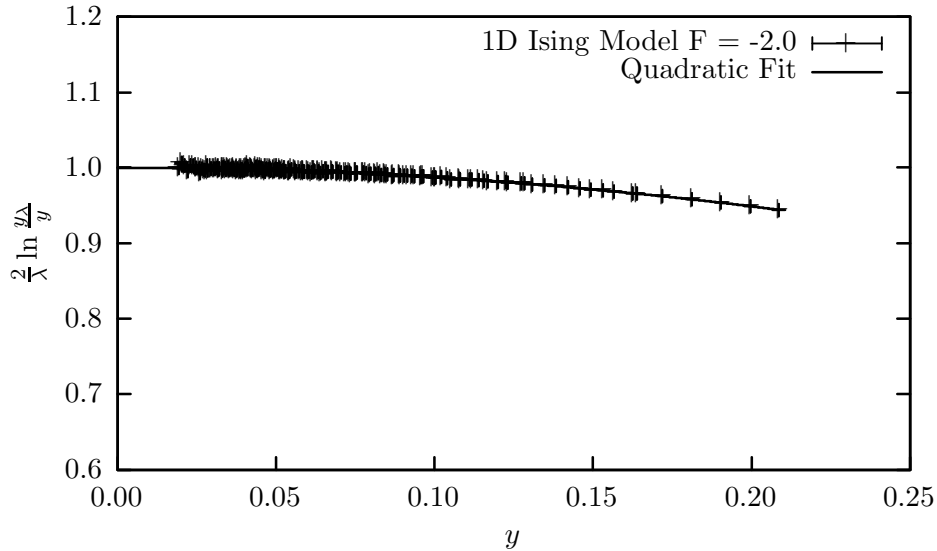}}
\caption{\label{Plot1d-I}Calculation of the critical exponent
  $\gamma_\lambda$. The crosses correspond to the numerical simulation
  data, whereas the line is a quadratic fit. The extrapolation to zero
  seems quite reliable, as the function is smooth enough. Errors are
  smaller than symbols.}
\end{figure}

\begin{figure}[h!]
\centerline{\input{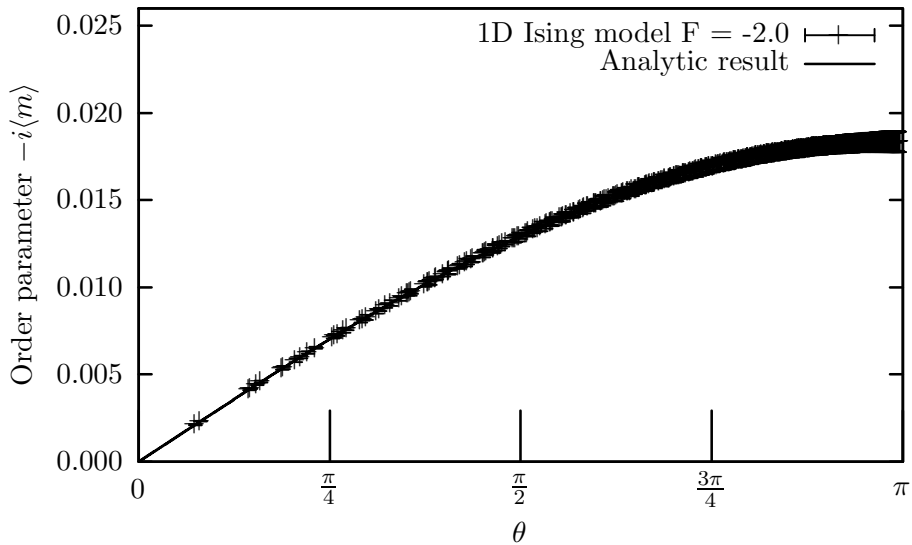}}
\caption{\label{Plot1d-II}Order parameter as a function of
  $\theta$. The non-zero value of the order parameter marks the
  spontaneous breaking of the $Z_2$ symmetry at $\theta=\pi$.}
\end{figure}

The first thing we did was to test the method in the one-dimensional
Ising model, and we checked the results against \eqref{Is-19}.  The
simulations were performed at a fixed volume, $N=1000$ spins, and
fixed reduced coupling $F=-2.0$. As the one-dimensional Ising model
has no phase transitions, and furthermore enjoys the exact scaling
property, there is no point in checking the method for several values
of the reduced coupling. The parameter we varied was the reduced
magnetic field $h$. As the simulations were quite fast, we could
obtain data for many values of $h$ with large statistics. In fact, for
each point in the plots we performed $10^7$ metropolis iterations. In
order to reduce autocorrelations, we performed at each iteration two
sweeps over the lattice, proposing metropolis changes in the
spins. The plots for the critical exponent and the order parameter are
shown in Figs. \ref{Plot1d-I} and \ref{Plot1d-II}.
Our result for the critical exponent from the fit in
fig. \ref{Plot1d-I} is
$$\gamma_\lambda = 0.99980 \pm 0.00008,$$ which agrees with the
analytical result. 

Then we simulated higher dimensional models, expecting to see
departures from this behaviour, as these models feature phase
transitions between ordered (antiferromagnetic) and disordered
phases. The two-dimensional simulations were done in a $100^2$
lattice, after $100000$ termalization sweeps. We spent $5000000$ steps
to measure each point accurately. The three-dimensional case, on the
other hand, used a $50^3$ volume, and measured each point for
$2500000$ steps after $100000$ steps of thermalization. The results
showed the expected departure in the behaviour. Our result for the
critical exponent $\gamma_\lambda \approx 2$ reveals a vanishing order
parameter at $\theta=\pi$ in the ordered phase ($F=-1.50$ for $2D$ and
$F=-1.00$ for $3D$), as shown in Figs. \ref{Plot2d-I}, \ref{Plot3d-I},
\ref{Plot2d-Ib} \footnote{Actually the $Z_2$ symmetry is spontaneously
  broken, for the staggered magnetization $m_S\neq0$
  \cite{LeeYang}. This point will be clarified in the mean-field
  approximation.}.
$$\gamma_\lambda^{2D} = 1.9997 \pm 0.0002$$
$$\gamma_\lambda^{3D} = 1.9998 \pm 0.0002$$ We can confirm this facts
by plotting the order parameter against $\theta$, as it is done in
Fig. \ref{Plot2d-II} and \ref{Plot3d-II}.
\begin{figure}[h!]
\centerline{\input{Pictures/Is2dPlot-I}}
\caption{\label{Plot2d-I}Calculation of the critical exponent
  $\gamma_\lambda$ in the ordered phase of the two-dimensional
  model. The pluses correspond to the numerical simulation data,
  whereas the line is a cuadratic fit. Errors are much smaller than
  symbols, except for the points lying close to the origin.}
\end{figure}
\begin{figure}[h!]
\centerline{\input{Pictures/Is3dPlot-I}}
\caption{\label{Plot3d-I}Calculation of the critical exponent
  $\gamma_\lambda$ in the ordered phase of the three-dimensional
  model. The pluses correspond to the numerical simulation data,
  whereas the line is a constant fit. Errors are much smaller than
  symbols, except for the points lying close to the origin.}
\end{figure}
\begin{figure}[h!]
\centerline{\input{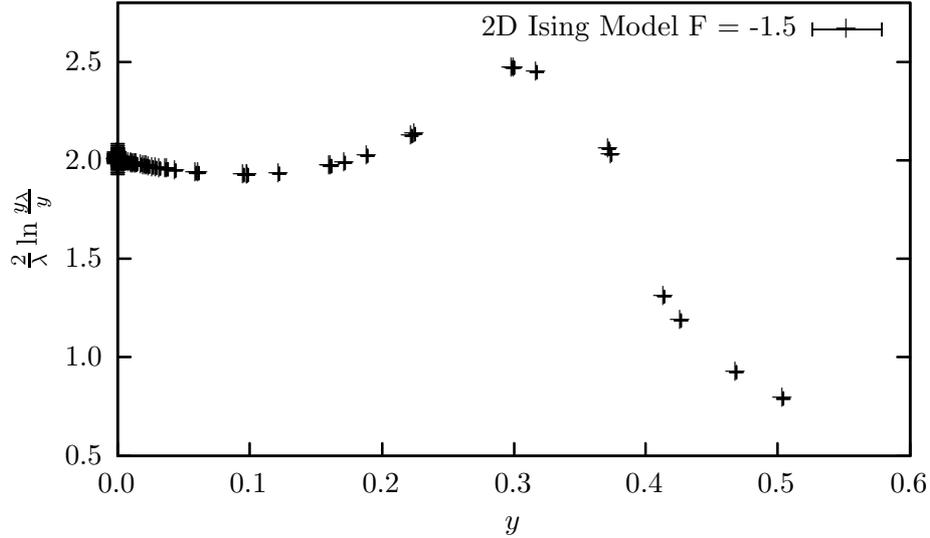}}
\caption{\label{Plot2d-Ib}Due to the peaked behaviour of the quotient
  $y_\lambda/y$ around $y=0.3$, the extrapolation to zero required
  many simulations at small values of the magnetic field. Errors are
  much smaller than symbols, except for the points lying close to the
  origin.}
\end{figure}
\begin{figure}[h!]
\centerline{\input{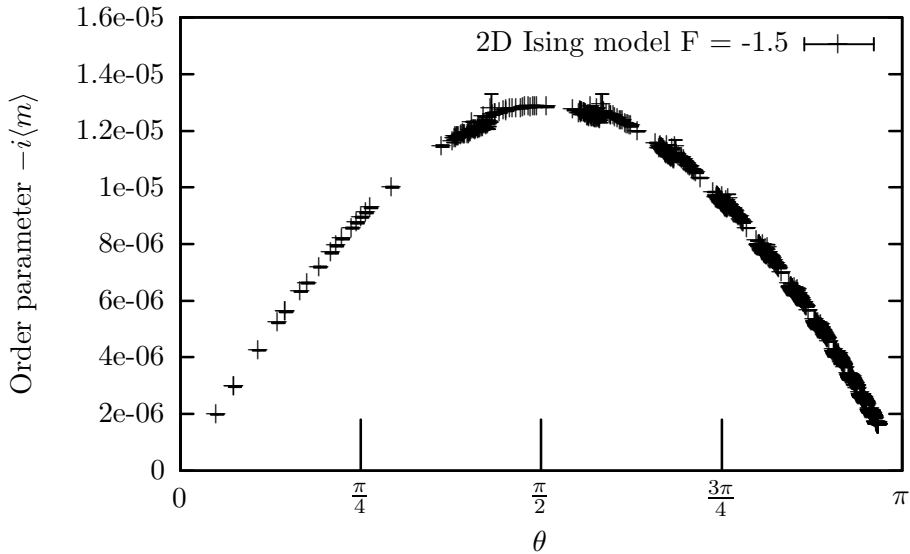}}
\caption{\label{Plot2d-II}Order parameter as a function of $\theta$ in
  the ordered phase of the two-dimensional model. It vanishes at
  $\theta=\pi$.}
\end{figure}
\begin{figure}[h!]
\centerline{\input{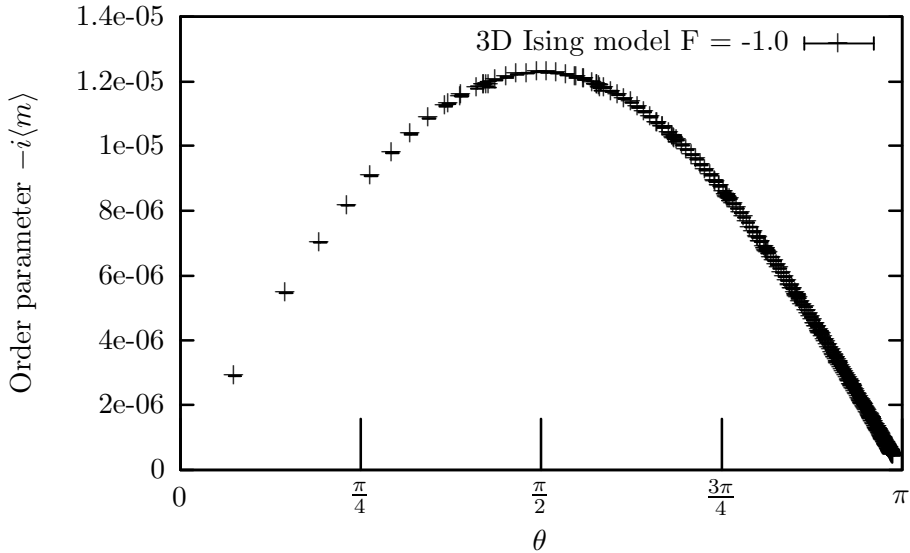}}
\caption{\label{Plot3d-II}Order parameter as a function of $\theta$ in
  the ordered phase of the three-dimensional model. As in its
  two-dimensional counterpart, it vanishes at $\theta=\pi$. Errors are
  smaller than symbols.}
\end{figure}

\begin{figure}[h!]
\centerline{\input{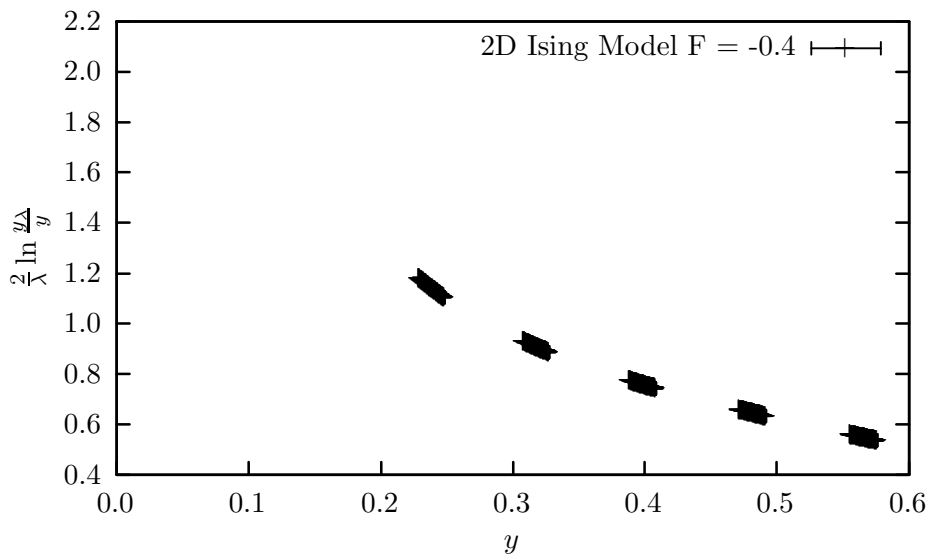}}
\caption{\label{Plot2d-III}Failed calculation of the critical exponent
  $\gamma_\lambda$ in the disordered phase for the two dimensional
  model. Our data is so far from the $y=0$ axis, that an extrapolation
  can not be used to find out the value of $\gamma_\lambda$. A peak
  for lower values of $y$, as the one appearing in
  Fig. \ref{Plot2d-Ib}, cannot be discarded `a priori'. Errors are
  much smaller than symbols.}
\end{figure}

\begin{figure}[h!]
\centerline{\input{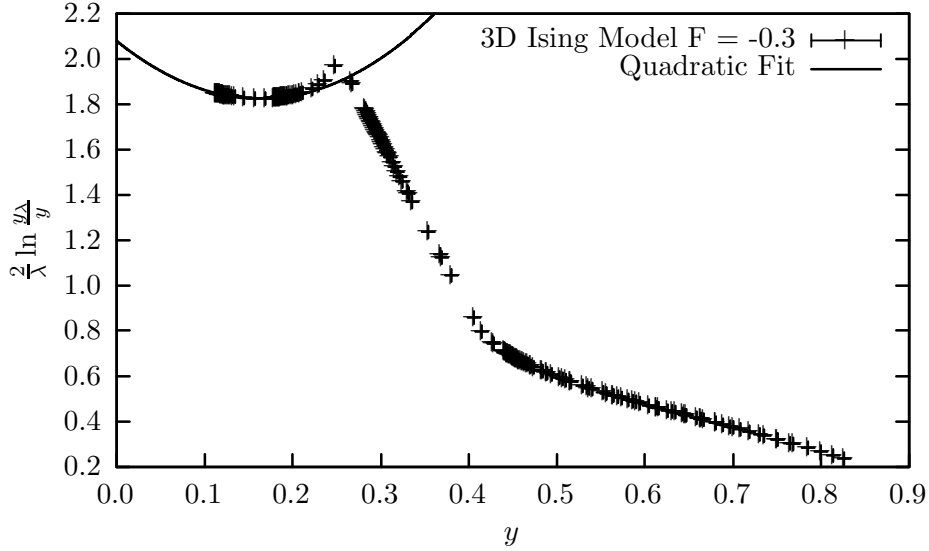}}
\caption{\label{Plot3d-III}Another calculation of the critical
  exponent $\gamma_\lambda$ in the ordered phase $F=-0.3$ for the
  three-dimensional model. Our data approaches the $y=0$ axis enough
  to try an extrapolation, but the result suffers from much larger
  errors than in the $F=-1.0$ case. Here
  $\gamma_\lambda=2.079\pm0.003$, and the measurement errors are much
  smaller than symbols.}
\end{figure}

The disordered phase revealed a caveat of this method, as it was
impossible for us to extrapolate the function
$\frac{y_\lambda}{y}\left(y\right)$ to zero. The reason is simple: at
small values of $F$, $y$ and $y_\lambda$ approach unity, for at
vanishing $F$ we recover the paramagnetic Langevin solution
$m=\tan\frac{\theta}{2}$ and $y=1$. The smaller the value of $F$, the
greater the gap between zero and our data becomes, and at some point,
the extrapolation is not reliable any more, and the results depend
strongly on the fitting function used. An example can be seen in
Fig. \ref{Plot2d-III}, where the data for the two-dimensional model at
$F=-0.40$ are plotted. In this case, we are too far from zero to find
out accurately the critical exponent, and the value of $F$ could not
be lowered much more, for the transition to the ordered phase is known
to happen at $F\sim-0.44$. In Fig. \ref{Plot3d-III} a similar example
is shown for the ordered phase in the three-dimensional model, but
this time a tentative extrapolation could be done, casting a reliable
result.

These examples show how this method works fine when the
antiferromagnetic couplings are strong enough. In general, as discussed at the
end of the previous section, the method 
performs well in the low temperature regime od the Ising model and for 
asymptotically free theories, whose continuum limit
lie in the region of weak coupling. In this region, the density of
topological structures is strongly suppressed. Thus the order
parameter and $y\left(z\right)$ take small values, making the plot
$\frac{y_\lambda}{y}\left(y\right)$ easily extrapolable to zero. In
the particular case of the antiferromagnetic Ising model, large values
of $\left|F\right|$ ensure a small magnetic susceptibility. A high value of the
dimension also helps, for instance, the three-dimensional model
requires a lower value of the coupling than the two-dimensional case
to make a reliable extrapolation of
$\frac{y_\lambda}{y}\left(y\right)$ to $y\rightarrow0$, for each spin
is affected by a higher number of neighbours.

\begin{figure}[h!]
\centerline{\input{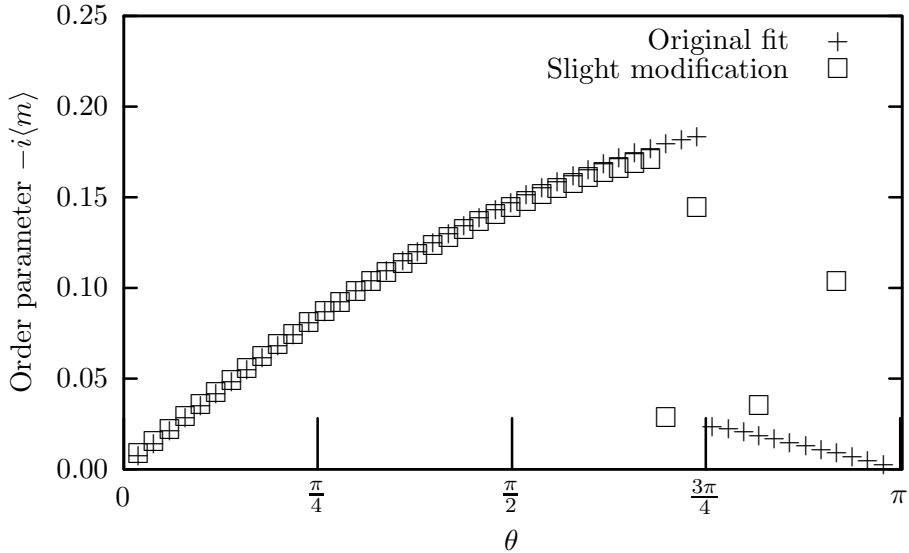}}
\caption{\label{Plot2d-IV}Failed calculation of the order parameter in
  the disordered phase using the improved p.d.f. method described at
  the beginning of this paper. In the first case, the points seem to
  predict a phase transition at $\theta\sim 2.35$, whereas in the
  second case the points depart sharply from a smooth function and
  never come back. The only difference between fits was the number of
  points used: in the second case, only half of the points (the
  closest to the origin) were used. Other modifications in the fitting
  procedures indicate us that the transition point is not stable. This
  might indicate either a failure in the fitting function, or a phase
  transition, and the impossibility for the method to precise the
  transition point, unless a perfect ansatz is made. Errors were not
  estimated.}
\end{figure}

As this method failed to deliver interesting results in the disordered
phase, we tried a different approach: we expected naively that the
two-dimensional model resemble the one-dimensional model at low values
of the coupling. Since the p.d.f. method worked well for the
one-dimensional case \cite{PdfIsing}, it made sense that we applied it
to the present scenario. What we found is an unstable behaviour:
sometimes the method seems to predict the phase transition, in the
sense that at finite volume there is not true phase transition, and an
abrupt modification in the order parameter, linking the two expected
behaviours, should happen. This is what we observe in one of the data
sets of Fig. \ref{Plot2d-IV}. Nonetheless, if a slightly different set
of points is taken to fit the saddle point equation \eqref{Is-19g},
the resulting data show a sharp departure from the expected behaviour
at some $\theta$.

There are two possible explanations to this behaviour: either the
fitting function selected is wrong, or there is some hidden phenomenon
we are overlooking. The fitting function used was an odd quotient of
polynomials $$\frac{ax^3 - x}{cx^2 - b}$$ which should account well
for the behaviour of the order parameter, given the assumption that it
is similar to the one-dimensional case. The addition of more terms to
the fit did not do much to improve the result, hence this possibility
was discarded.

The existence of a phase transition in the middle, however, was an
interesting option. Indeed, the two-dimensional model in the presence
of a $\theta$ term was solved exactly at the point $\theta=\pi$ almost
sixty years ago by Yang and Lee in \cite{LeeYang}, and reviewed again
in \cite{Shrock}. In those papers, a phase diagram was proposed were
the antiferromagnetic model always stayed in an ordered phase at any
non-zero value of $F$. Since the system is in a disordered state for
low $F$'s and zero field, some phase transition has to occur in the
middle. Thence, the failure of the p.d.f. method should be due to a
poor ansatz for the fitting function, caused by the presence of a
phase transition at some $\theta_c$.

The fact that the results for the two- and three-dimensional models
are qualitatively the same in the ordered phase, makes us wonder
whether this behaviour changes for some value of the dimension
$D>3$. Moreover, the behaviour of this model in the disordered phase
is unknown to us. That is why we decided to carry out a mean-field
approximation of the model, and compute the critical exponent
$\gamma_\lambda$. As we know, mean-field results for other critical
exponents are exact for the $n$-dimensional ferromagnetic Ising model,
provided that $n\geq 4$. Thus we expect that, if the mean-field result
for $\gamma_\lambda$ is the same to that of the two- and the
three-dimensional Ising model, then $\gamma_\lambda = 2$ for any value
of the dimension.

\section{Mean-field calculation}

In antiferromagnetic compounds the spin-alignment pattern is
staggered, hence, in order to define the mean-field version of the
antiferromagnetic Ising model, we should divide the lattice in two
sublattices, and define a coupling among spins whose sign depends on
whether these two spins are on the same sublattice or not. For spins
belonging to the same lattice, the coupling should be ferromagnetic
($J>0$), but for spins belonging to different sublattices, the
coupling should favour antiparallel ordering ($-J<0$), according to
the antiferromagnetic nature of the system. Therefore, two different
mean-fields should appear, $\langle s_i\rangle_{i\in S_1} = m_1$ and
$\langle s_i\rangle_{i\in S_2} = m_2$, referring to each one of the
different sublattices. 

A reasonable mean field hamiltonian compatible with these requirements
is \footnote{Of course other mean-field approaches could be used, but
  we expect to obtain the same qualitative picture for the phase
  diagram. For example, a calculation in the scheme of
  \cite{AntiferIs} produces similar results, and more importantly the
  same value for $\gamma_\lambda$ as the one in our calculation.}
\begin{equation}
H\left(J,B,\left\{s_i\right\}\right) = -\frac{J}{N}\left(\sum_{i\in S_1}s_i - \sum_{j\in S_2}s_j\right)^2
- B_1\sum_{i\in S_1}s_i - B_2\sum_{j\in S_2}s_j.
\label{Is-37b}
\end{equation}

We define as before $h_i = \frac{2 B_i}{KT}$ (for each sublattice $i=1,2$) and $F = \frac{J}{KT}$, and
the usual and the staggered magnetizations, $m = m_1 + m_2$, $m_S =
m_1 - m_2$. We are interested in the model for imaginary values of the
field, $h \rightarrow i\theta$. Using standard saddle point techniques
(see appendix A for details) we obtain the mean field equations:
\begin{eqnarray}
\label{Is-47c}m\, =& \frac{1}{2}\frac{i\sin\theta}
{\cosh^2\left(2\left|F\right|m_S\right) - \sin^2\frac{\theta}{2}},\\
\label{Is-47d}m_S =& \frac{1}{2}\frac{\sinh\left(4\left|F\right|m_S\right)}
{\cosh^2\left(2\left|F\right|m_S\right) - \sin^2\frac{\theta}{2}}.
\end{eqnarray}

\begin{figure}[h!]
\centerline{\input{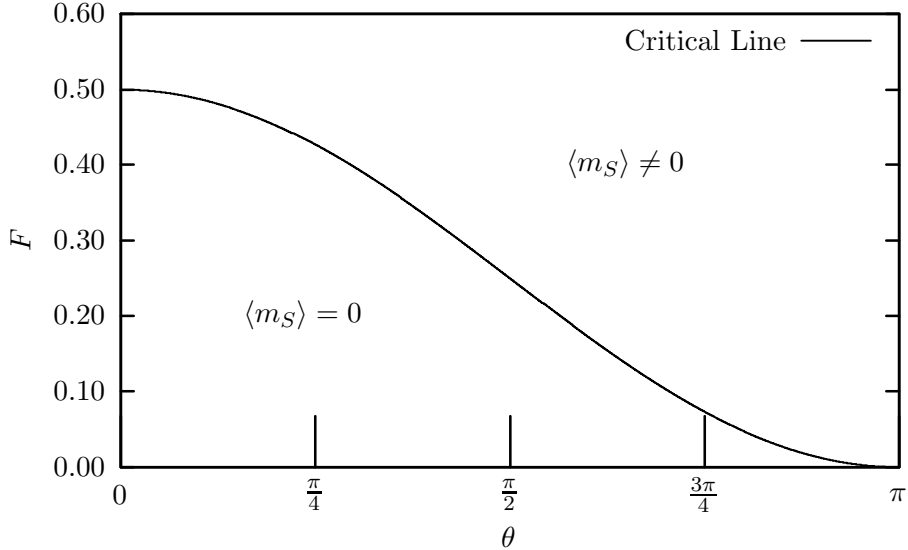}}
\caption{\label{PhaseD}Phase diagram of the mean-field approximation
  to the antiferromagnetic Ising model in the $F-\theta$ plane.}
\end{figure}

An analysis of these equations shows that for low values of $F$
there's only one solution, $m_S = 0$, corresponding to a paramagnetic
phase. For values over a certain $F_c$ the situation changes and two
new symmetric solutions appear, which are in fact the physically
relevant ones.  From the saddle point equations, \eqref{Is-46} and
\eqref{Is-46a}, $F_c$ can be obtained as a function of $\theta$:
\begin{equation}
2F_c = \cos^2\frac{\theta_c}{2}.
\label{Is-48a}
\end{equation}
The phase diagram of the system in the F-$\theta$ plane is shown in
Fig. \ref{PhaseD},
There is a second order phase transition at the critical line
\eqref{Is-48a}. As $\theta\rightarrow\pi$ the paramagnetic phase
narrows, until it is reduced to the single point $F=0$ at
$\theta=\pi$. The staggered phase (with antiparallel spin ordering
$m_S\neq0$ and $F>\frac{\cos^2\frac{\theta}{2}}{2}$) on the other hand
features $Z_2$ spontaneous symmetry breaking at $\theta=\pi$, as
equation \ref{Is-47d} shows. The fact that this model features a phase
transition at non-zero values of the external field is quite
remarkable. This kind of transitions would never appear in a
ferromagnetic model within a real external magnetic field, as the
external field and the spin coupling work in the same direction:
parallel spin alignment. On the contrary, in the antiferromagnetic
case, the introduction of a real external field produces frustration,
whose origin comes from the competition of the spin coupling, trying
to move the spins towards an antiparallel configuration, and the
external field, favouring a completely parallel structure. What is
remarkable of these results is the fact that in the antiferromagnetic
case an imaginary magnetic field, strong enough, is able to move the
system from the paramagnetic phase to a phase with antipararell
ordering.

All the magnetizations are continuous functions, but the staggered
susceptibility $\chi_S$ diverges, as usually happens in a second order
phase transition.  The \emph{topological}\footnote{Topological in the
  sense that $M/2$ is a quantized charge, $m/2$ is its associated
  charge density, and $\chi_T$ the susceptibility.} susceptibility
$\chi_T$, on the other hand, displays a gap at the critical line. The
computation (see Appendix A) gives the result
\begin{equation}
\Delta\chi_T = \lim_{\theta\rightarrow\theta_c^+}\chi_T -
\lim_{\theta\rightarrow\theta_c^-}\chi_T =
\frac{3}{4\left|F\right|}\frac{2\left|F\right|-1}{4\left|F\right|-3}.
\label{Is-48b}
\end{equation}

Finally, the critical exponent $\gamma_\lambda$ for this mean-field
theory can be calculated, to see if it coincides with that obtained in
simulations. In order to do so, we expand $m$ in the neighbourhood of
$\theta=\pi$
\begin{equation}
m\left(\theta\right) \sim m\left(\pi\right) + \left.\frac{\partial
  m}{\partial \theta}\right|_{\theta=\pi}\left(\pi-\theta\right) +
\left.\frac{\partial^2 m}{\partial
  \theta^2}\right|_{\theta=\pi}\left(\pi-\theta\right)^2 + \ldots
\label{Is-49}
\end{equation}
If $\gamma_\lambda$ is not natural number, we expect the first
non-zero derivative to diverge. On the contrary, if $\gamma_\lambda$
is a natural number, the order of the first non-vanishing derivative
will give us the critical exponent. Taking derivatives
\begin{equation}
\left.\frac{\partial m}{\partial\theta}\right|_{\theta=\pi} =
\frac{i}{2}\frac{\left(2\cos^2\theta -
  1\right)\cosh^2\left(2\left|F\right|m_S\right) + 1 -
  \cos^2\frac{\theta}{2}}
     {\left(\cosh^2\left(2\left|F\right|m_S\right)-\sin^2\frac{\theta}{2}\right)^2}
     -$$
$$-\left.\frac{i}{2} \frac{2\left|F\right|\sin\theta
       \sinh\left(4\left|F\right|m_S\right)\left.  \frac{d
         m_S}{d\theta}\right|_{\theta=\pi}}
     {\left(\cosh^2\left(2\left|F\right|m_S\right)-\sin^2\frac{\theta}{2}\right)^2}
     \right|_{\theta=\pi} =$$
$$= -\frac{i}{2}\left[\frac{1}{\sinh^2\left(2\left|F\right|m_S\right)}
       +
       \frac{2\left|F\right|\sin\theta\sinh\left(4\left|F\right|m_S\right)\left.\frac{d
           m_S}{d\theta}\right|_{\theta=\pi}}{\sinh^4\left(2\left|F\right|m_S\right)}\right]
\label{Is-50}
\end{equation}
The first term on the r.h.s. of \eqref{Is-50} does not diverge since
$m_S$ is not vanishing as $\theta\rightarrow\pi$. The second term is
proportional to
$$\lim_{\theta\rightarrow\pi}\frac{d m_S}{d\theta}\sin\theta.$$ After
a tedious calculation, it can be shown that it vanishes, therefore
\begin{equation}
m\left(\theta\right) \sim
i\frac{\pi-\theta}{2\sinh^2\left(2\left|F\right|m_s\right)} \sim
K\left(\pi-\theta\right),
\label{Is-51}
\end{equation}
with $K$ a non-zero constant. The magnetization behaves as
$\left(\pi-\theta\right)^{\gamma_\lambda-1}$ in the neighbourhood of
$\theta=\pi$. Thence, for mean-field antiferromagnetic theory
$\gamma_\lambda = 2$ for $F\neq 0$. For $F=0$ $\gamma_\lambda = 0$ and
the magnetization diverges as $\tan\frac{\theta}{2}$ when
$\theta\rightarrow\pi$.

Since mean-field theory works better in high dimensional systems (it
reproduces all the critical exponents exactly for the ferromagnetic
Ising model in dimension 4 and above), and the exponent
$\gamma_\lambda$ seems to have settled in $\gamma_\lambda = 2$ for the
two- and three-dimensional models, and for the mean-field
approximation, we expect this result to hold for any dimension of the
system. This is not a proof, but in fact, it would be very remarkable
if the behaviour of the antiferromagnetic Ising model in a higher
dimension departed from $\gamma_\lambda = 2$.

\section{Conclusions}

Although the aim behind this investigation of the antiferromagnetic
Ising model is to test our techniques for a future application to QCD
in presence of a $\theta$ term, the results obtained through this work
deserve attention on their own merit. Using the method described in
section \ref{MiIs}, the order parameter for the $Z_2$ symmetry can be
calculated for any value of $\theta$, and although there are some
regions of the phase diagram where the method does not work well (high
temperature regime), it provided us with enough information to make an
educated guess on the phase diagram of the theory.

Our guess was later confirmed by a mean-field calculation, which
shares many properties with the original model. The results of
\cite{LeeYang} and \cite{Shrock} supplied the remaining information
for the two-dimensional case. In the end, we were able to reconstruct
qualitatively the whole phase diagram of the theory for
two-dimensions, and although we did not pursue to solve the model for
higher dimensions, the mean-field calculations, and the fact that the
behaviour for the two- and the three-dimensional models is the same
for large values of $F$, give us strong indications that this phase
diagram is qualitatively valid for any dimension of the model larger
than one. The one-dimensional model is an exceptional case in which
only one phase appears with spontaneous magnetization at
$\theta=\pi$. On the contrary for $d=2,3$ our numerical simulations
show a density of magnetization that continuously vanishes at
$\theta=\pi$ at low temperatures, and the mean field calculation
strongly suggests that this result holds for any temperature and any
larger dimension. However this does not mean that the $Z_2$ symmetry
of the model at $\theta=\pi$ is realized in the ground state since the
mean field calculation shows a non vanishing value of the staggered
magnetization at $\theta=\pi$. Indeed there is a region of
non-vanishing measure in the $F-\theta$ plane, including the
$\theta=\pi$ line, where the staggered magnetization does not vanish
and the saddle point equation \eqref{Is-47d} shows two symmetric
solutions for this quantity. In all this phase translational invariance
is spontaneously broken and in the $\theta=\pi$ line parity is also
spontaneously broken.

The method only has two caveats: 
\begin{enumerate}[i.]
\item 
It does not work properly (and can give wrong results) if there is a
phase transition for some $\theta<\pi$.
\item 
For small absolute values of the coupling $F$ (high temperatures) the
required extrapolations are not feasible. 
\end{enumerate}

Fortunately, the standard wisdom on QCD, based on reasonable
assumptions, expects only one phase transitions at $\theta=\pi$
\cite{ettore}, and QCD is an asymptotically free theory, thus its
continuum limit lies, as discussed in section 3, in the ``low temperature''
regime 
where our approach works well. Therefore
this method has become the perfect
candidate to perform simulations of QCD with a $\theta$ term, which
might provide precious information concerning the strong $CP$ problem.

A final remark on the antiferromagnetic Ising model at $\theta=\pi$ is
pertinent. By using polymerization techniques an algorithm able to
simulate the model at $\theta=\pi$ and free from the sign problem can
be developed, and it could be useful to test if the mean field
predictions reported here are verified for any dimension larger than 1
and any temperature at $\theta=\pi$.

\section{Acknowledgments} 
This work has been partially supported by MICINN (grant FPA2009-09638 and grant FPA2008-01732),
DGIID-DGA (grant2007-E24/2) and by the European Union under Grant
Agreement number PITN-GA-2009-238353 (ITN STRONGnet).  E. Follana is
supported by MICINN through the Ram\'on y Cajal program, and
A. Vaquero was supported by MICINN through the FPU program.

\appendix

\section{Mean field model}

The hamiltonian of the mean-field model is:
\begin{equation}
H\left(J,B,\left\{s_i\right\}\right) = -\frac{J}{N}\left(\sum_{i\in S_1}s_i - \sum_{j\in S_2}s_j\right)^2
- B_1\sum_{i\in S_1}s_i - B_2\sum_{j\in S_2}s_j.
\label{Is-37ap}
\end{equation}
where we have assummed that an independent field $B_i$ acts on each sublattice $i=1,2$. This modification will allow us to compute separately $m_1$ and $m_2$ once we have obtained the free energy; but of course, for the case of an uniform external field we shuld set $B = B_1 = B_2$ at the end of the calculation.

We define as before $h_i = \frac{2 B_i}{KT},\quad i=1,2$ and $F = \frac{J}{KT}$, and
the usual and the staggered magnetizations, $m = m_1 + m_2$, $m_S =
m_1 - m_2$. The partition function
\begin{equation}
Z\left(F,h\right) = \sum_{\{s_i\}} e^{\frac{F}{N}\left({\sum \atop
    {i\in S_1}}s_i - {\sum \atop {j\in S_2}}s_j\right)^2 +
  \frac{h_1}{2} {\sum \atop {i\in S_1}}s_i + \frac{h_2}{2} {\sum \atop
    {j\in S_2}}s_j}
\label{Is-39}
\end{equation}
can be summed up by applying the Hubbard-Stratonovich identity to
linearize the exponent
\begin{equation}
Z\left(F,h\right) = \frac{1}{\pi^\frac{1}{2}}\int_{-\infty}^{\infty}
\sum_{\{s_i\}} e^{-x^2 +
  \left[2x\frac{\left|F\right|^\frac{1}{2}}{N^\frac{1}{2}} +
    h_1\right]{\sum \atop {i\in S_1}} s_i -
  \left[2x\frac{\left|F\right|^\frac{1}{2}}{N^\frac{1}{2}} - h_2\right]
       {\sum \atop {j\in S_2}} s_j}dx.
\label{Is-41}
\end{equation}
At this point we see that the introduction of the $\theta$ term
through the transformations
$$h_1\rightarrow i\theta_1,\qquad h_2\rightarrow i\theta_2,$$ render
the hyperbolic cosines complex. The $\frac{1}{2}$ factor allows us to
define properly the quantized number $\frac{M}{2}$. The integrand
factorizes, as there is no spin-spin interaction

\begin{equation}
Z\left(F,h\right) =
\frac{2^N}{\pi^{\frac{1}{2}}}\int_{-\infty}^{\infty} e^{-x^2}
\left[\cosh\left(2x\frac{\left|F\right|^\frac{1}{2}}{N^\frac{1}{2}} +
  i\frac{\theta_1}{2}\right)\right.\times$$
$$\left.\cosh\left(2x\frac{\left|F\right|^\frac{1}{2}}{N^\frac{1}{2}}
  - i\frac{\theta_2}{2}\right)\right]^\frac{N}{2}dx.
\label{Is-42}
\end{equation}
Now we bring the transformation
$$\begin{array}{ccc} x & \rightarrow & N^{\frac{1}{2}}y\\ dx&
  \rightarrow & N^{\frac{1}{2}}dy
\end{array}$$
so \eqref{Is-42} becomes

\begin{equation}
Z\left(F,h\right) = \frac{2^N
  N^\frac{1}{2}}{\pi^\frac{1}{2}}\int_{-\infty}^{\infty}
\left[e^{-y^2 +}\right.$$
$$\left.^{\frac{1}{2}\ln\left[\cosh\left(2\left|F\right|^\frac{1}{2}y
      +
      i\frac{\theta_1}{2}\right)\cosh\left(2\left|F\right|^\frac{1}{2}y
      - i\frac{\theta_2}{2}\right)\right]}\right]^Ndy.
\label{Is-43}
\end{equation}
where we have written the whole integral as an exponential. We can
evaluate the integral in the large $N$ limit using the
\emph{saddle-point technique}.
\begin{equation}
\lim_{N\rightarrow\infty}\frac{1}{N}\ln Z\left(J,B\right) = \ln 2 + $$
$$ + \lim_{N\rightarrow\infty}\frac{1}{N}\ln\int_{-\infty}^{\infty}
\left[e^{-y^2 +
    \frac{1}{2}\ln\left[\cosh^2\left(2\left|F\right|^\frac{1}{2}y\right)
      - \sin^2\frac{\theta}{2}\right]}\right]^Ndy.
\label{Is-44}
\end{equation}
The maximum of
\begin{equation}
g\left(y\right) = -y^2 +
\frac{1}{2}\ln\left[\cosh^2\left(2\left|F\right|^\frac{1}{2}y\right) -
  \sin^2\frac{\theta}{2}\right]
\label{Is-45}
\end{equation}
gives us the saddle-point equations
\begin{eqnarray}
\label{Is-46} -y_0 + \frac{\left|F\right|^\frac{1}{2}}{2}
\frac{\sinh\left(4\left|F\right|^\frac{1}{2}y_0\right)}
{\cosh^2\left(2\left|F\right|^\frac{1}{2}y_0\right) - \sin^2\frac{\theta}{2}} = 0,\\
\label{Is-46a}-1 + 2\left|F\right|\frac{\cos^2
\frac{\theta}{2}\cosh\left(4\left|F\right|^\frac{1}{2}y_0\right) -
\sinh^2\left(2\left|F\right|^\frac{1}{2}y_0\right)}
      {\cosh^2\left(2\left|F\right|^\frac{1}{2}y_0\right) -
        \sin^2\frac{\theta}{2}} < 0.
\end{eqnarray}
Thus, the free energy is
\begin{equation}
f\left(F,h\right) = \ln2 + g\left(y_0\right)
\label{Is-46b}
\end{equation}
where $y_0$ verifies \eqref{Is-46}, \eqref{Is-46a}.

We can also evaluate the magnetizations:
\begin{eqnarray}
\label{Is-46c}m_1 =& \frac{1}{2}\frac{\cosh\left(2\left|F\right|^\frac{1}{2}y_0\right)
\sinh\left(2\left|F\right|^\frac{1}{2}y_0\right) 
+ i\sin\frac{\theta}{2}\cos\frac{\theta}{2}}
{\cosh^2\left(2\left|F\right|^\frac{1}{2}y_0\right) - \sin^2\frac{\theta}{2}},\\
\label{Is-46d}m_2 =& -\frac{1}{2}\frac{\cosh\left(2\left|F\right|^\frac{1}{2}y_0
\right)\sinh\left(2\left|F\right|^\frac{1}{2}y_0\right) 
- i\sin\frac{\theta}{2}\cos\frac{\theta}{2}}
{\cosh^2\left(2\left|F\right|^\frac{1}{2}y_0\right) - \sin^2\frac{\theta}{2}},\\
\label{Is-46e}m =& \frac{i\sin\frac{\theta}{2}\cos\frac{\theta}{2}}
{\cosh^2\left(2\left|F\right|^\frac{1}{2}y_0\right) - \sin^2\frac{\theta}{2}},\\
\label{Is-46f}m_S =& \frac{\cosh\left(2\left|F\right|^\frac{1}{2}y_0\right)
\sinh\left(2\left|F\right|^\frac{1}{2}y_0\right)}
      {\cosh^2\left(2\left|F\right|^\frac{1}{2}y_0\right) -
        \sin^2\frac{\theta}{2}}.
\end{eqnarray}
Therefore, and using \eqref{Is-46},
\begin{equation}
y_0 = \left|F\right|^\frac{1}{2}\langle m_S\rangle.
\label{Is-47}
\end{equation}

In order to compute the gap in the susceptibilities at the critical
line, we must examine the behaviour of the variable $y_0$ in the
neighbourhood of $\theta_c$.  The way to proceed is to expand the
hyperbolic functions in (\ref{Is-46}) as a power series in $y_0$. For
$\theta<\theta_c$ the only solution to the saddle point equation is
$y_0=0$, so we expand around this point
\begin{eqnarray}
\label{AppX-IIa}\sinh\left(4\left|F\right|^\frac{1}{2}y_0\right) 
= 4\left|F\right|^\frac{1}{2}y_0 +
\frac{32}{3}\left|F\right|^\frac{3}{2}y_0^3 + O\left(y^5\right),\\
\label{AppX-IIb}\cosh\left(2\left|F\right|^\frac{1}{2}y_0\right) 
= 1 + 2\left|F\right|y_0^2 + O\left(y^4\right),
\end{eqnarray}
so the saddle-point equation becomes
\begin{equation}
y_0 \sim 2\left|F\right|\frac{y_0 + \frac{8}{3}\left|F\right|y_0^3}
{4\left|F\right|y_0^2 + \cos^2\frac{\theta}{2}}\qquad y_0<<1.
\label{AppX-III}
\end{equation}
Now we expand again the denominator of \eqref{AppX-III} up to $y_0^2$,
\begin{equation}
\frac{1}{4\left|F\right|y_0^2 + \cos^2\frac{\theta}{2}} =
\frac{1}{\cos^2\frac{\theta}{2}} -
\frac{4\left|F\right|}{\cos^4\frac{\theta}{2}}y_0^2 +
O\left(y_0^4\right).
\label{AppX-IV}
\end{equation}
Therefore, for $\theta\gsim\theta_c$,
\begin{equation}
y_0\sim\frac{2\left|F\right|}{\cos^2\frac{\theta}{2}}y_0 +
8\left|F\right|^2\frac{2\cos^2\frac{\theta}{2} -
  3}{3\cos^4\frac{\theta}{2}}y_0^3.
\label{AppX-V}
\end{equation}
We already know of the $y_0=0$ solution. Solving the quadratic equation that is left,
\begin{equation}
y_0 =
\sqrt{\frac{3}{8\left|F\right|^2}\frac{\left(\cos^2\frac{\theta}{2} -
    2\left|F\right|\right)\cos^2\frac{\theta}{2}}{2\cos^2\frac{\theta}{2}
    - 3}}.
\label{AppX-VI}
\end{equation}
Thus $y_0$ tends to zero as $\theta$ approaches the critical value for
a given $F$. Its derivative with respect to $\theta$, on the other
hand,
\begin{equation}
\frac{d y_0}{d\theta} = -\sqrt{\frac{3}{32\left|F\right|^2}}
\frac{\sin\theta\left(\cos^4\frac{\theta}{2}-3\cos^2\frac{\theta}{2}+3\left|F\right|\right)}
     {\sqrt{\left(2\cos^2\frac{\theta}{2} - 3\right)^3
         \left[\left(\cos^2\frac{\theta}{2} -
           2\left|F\right|\right)\cos^2\frac{\theta}{2}\right]}},
\label{AppX-VII}
\end{equation}
diverges as
$$\frac{1}{\sqrt{\left(\cos^2\frac{\theta}{2} -
    2\left|F\right|\right)\cos^2\frac{\theta}{2}}},$$ for at the
critical line $\cos\frac{\theta_c}{2} = 2F$. The divergence cancels in
the product $y_0\frac{dy_0}{d\theta}$. As $y_0 =
\sqrt{\left|F\right|}m_S$, this also applies to
$m_S\frac{dm_S}{d\theta}$.

The solution obtained in \eqref{AppX-VI} can be used to calculate the
behaviour of the susceptibilities around the critical point. The
`topological' susceptibility
\begin{equation}
\chi_T = \frac{d m}{d i\frac{\theta}{2}} = \frac{d m}{d\theta}
\frac{d\theta}{d i\frac{\theta}{2}} =
\frac{\left(\cos^2\frac{\theta}{2}-1\right)
  \cosh\left(2\left|F\right|m_S\right)+1-\cos^2\frac{\theta}{2}}
     {\left(\cosh^2\left(2\left|F\right|m_S\left(\theta\right)\right)-\sin^2
       \frac{\theta}{2}\right)^2}
     -$$ $$-\frac{2\left|F\right|\sin\theta\sinh
       \left(4\left|F\right|m_S\right)\frac{d m_S}{d\theta}}
     {\left(\cosh^2\left(2\left|F\right|m_S\right)
       -\sin^2\frac{\theta}{2}\right)^2},
\label{AppX-VIII}
\end{equation}
takes the value
\begin{equation}
\lim_{\theta\rightarrow\theta^-_c}\chi_T =
\frac{1}{\cos^2\frac{\theta_c}{2}} = \frac{1}{2\left|F\right|},
\label{AppX-IX}
\end{equation}
as we approach $\theta_c$ from below. However, if we come from the
antiferromagnetic phase $\theta>\theta_c$, the second term gives a
non-zero contribution, for the derivative $\frac{d m_S}{d\theta}$
diverges at the critical line. The divergence is cancelled exactly by
the factor $\sinh\left(4\left|F\right|m_S\right)$, as explained
before, and what remains is a finite contribution
$$\left.m_S\frac{d m_S}{d\theta}\right|_{\theta=\theta_c} =
-\frac{3\sin\theta_c}{16\left|F\right|^2\left(4\left|F\right| -
  3\right)}$$
$$\frac{2\left|F\right|\sin\theta\sinh\left(4\left|F\right|m_S\right)\frac{d
    m_S}{d\theta}}{\left(\cosh^2\left(2\left|F\right|m_S\right)
  -\sin^2\frac{\theta}{2}\right)^2}\sim-\frac{3\sin^2\theta_c}
{32\left|F\right|^2\left(4F-3\right)}$$. 
In the end
\begin{equation}
\lim_{\theta\rightarrow\theta^+_c}\chi_T = \frac{1}{2\left|F\right|} +
\frac{3}{4\left|F\right|}\frac{2\left|F\right|-1}{4\left|F\right|-3},
\label{AppX-X}
\end{equation}
and the gap is
\begin{equation}
\Delta\chi_T = \lim_{\theta\rightarrow\theta^+_c}\chi_T -
\lim_{\theta\rightarrow\theta^-_c}\chi_T =
\frac{3}{4\left|F\right|}\frac{2\left|F\right|-1}{4\left|F\right|-3}.
\label{AppX-XI}
\end{equation}

The staggered susceptibility diverges at the critical line. This is
quite expected, as for $\theta=0$ the susceptibility diverges at the
critical point. In order to obtain $\chi_S$, we need to take
derivatives with respect to a staggered field $\theta_S$, and then
take the $\theta_S\rightarrow0$ limit. To this purpose, we use the
original form of the free energy \eqref{Is-43} with a $\theta_S$ term
\begin{equation}
f\left(F,m_S,\theta,\theta_S\right) = -\left|F\right|m_S^2 +
\frac{1}{2} \ln\left[\cosh\left(2\left|F\right|m_S + \frac{i\theta +
    i\theta_S}{2}\right)\times\right.$$
$$\left.\cosh\left(2\left|F\right|m_S - \frac{i\theta - i\theta_S}{2}\right)\right]
\label{AppX-XII}
\end{equation}
Taking derivatives with respect to $m_S$ we should recover the
saddle-point equation with the addition of the $\theta_S$ source
\begin{equation}
\frac{df}{d m_S} = 0 = -2\left|F\right|m_S + \left|F\right|
\left[\tanh\left(2\left|F\right|m_S + \frac{i\theta +
    i\theta_S}{2}\right) +\right.$$
$$\left.\tanh\left(2\left|F\right|m_S - \frac{i\theta - i\theta_S}{2}\right)\right].
\label{AppX-XIII}
\end{equation}
From \eqref{AppX-XIII} a new equation for the staggered magnetization
is obtained
\begin{equation}
m_S = \frac{1}{2}\left[\tanh\left(2\left|F\right|m_S + \frac{i\theta +
    i\theta_S}{2}\right) +\right.$$
$$\left.\tanh\left(2\left|F\right|m_S - \frac{i\theta - i\theta_S}{2}\right)\right].
\label{AppX-XIVa}
\end{equation}
The derivative with respect to $\theta_S$ gives us the susceptibility
\begin{equation}
\chi_S = \frac{dm_S}{d\frac{i\theta_S}{2}} = \frac{1+2\left|F\right|\chi_S}
{2\cosh^2\left(2\left|F\right|m_S + \frac{i\theta + i\theta_S}{2}\right)} +$$
$$\frac{1+2\left|F\right|\chi_S}{2\cosh^2\left(2\left|F\right|m_S -
  \frac{i\theta - i\theta_S}{2}\right)} =
\frac{1+2\left|F\right|\chi_S}{2}X,
\label{AppX-XIVb}
\end{equation}
where
$$X = \frac{1}{2\cosh^2\left(2\left|F\right|m_S + \frac{i\theta +
    i\theta_S}{2}\right)} + \frac{1}{2\cosh^2\left(2\left|F\right|m_S
  - \frac{i\theta - i\theta_S}{2}\right)}.$$ Moving all the terms
proportional to $\chi_S$ to the l.h.s.
\begin{equation}
2\chi_S\left(1-\left|F\right|X\right) = X,
\label{AppX-XV}
\end{equation}
we can find the value of $\chi_S$
\begin{equation}
\chi_S = \frac{X}{2-2\left|F\right|X}.
\label{AppX-XVI}
\end{equation}
The quantity $X$ must be evaluated at the point $\theta=\theta_c$ and
$\theta_S=0$. This is not a difficult task and the final value is
$$X = \frac{1}{\left|F\right|}.$$ The staggered susceptibility, on the
other hand, diverges at the critical line. Approaching the critical
point from the high temperature region (where the only solution to the
saddle point equation is $m_S = 0$), we find
\begin{equation}
\chi_S = \frac{1}{2\left|F_c\right|-2\left|F\right|},
\label{AppX-XVII}
\end{equation}
the susceptibility diverges at the critical line $F=F_c$ and the
critical exponent for this divergence
\begin{equation}
\chi_S \propto \left|T-T_c\right|^{-\gamma_S}
\label{AppX-XVIIb}
\end{equation}
is $\gamma_S=1$.

Finally, and to elucidate the behaviour of $m\left(\theta\right)$ as
$\theta\rightarrow\pi$, we need to work out the following limit
\begin{equation}
\lim_{\theta\rightarrow\pi}\frac{dm_S}{d\theta}\sin\theta.
\label{AppX-XVIII}
\end{equation}
As $\sin\theta\rightarrow0$ when $\theta$ approaches $\pi$, only if
the derivative $\frac{dm_S}{d\theta}$ diverges at $\theta=\pi$ is the
product \eqref{AppX-XVIII} non-vanishing. The expansion we performed
previously is not very useful here, as the point $\theta=\pi$ is far
from the critical line (unless we are taking the $F\rightarrow0$ limit
as well). The way to solve this problem is to compute implicitly the
derivative from the saddle-point equation \eqref{Is-47d} at
$\theta=\pi$
\begin{equation}
\left.\frac{dm_S}{d\theta}\right|_{\theta=\pi} = 
\left.\frac{dm_S}{d\theta}\frac{2\left|F\right|\cosh\left(4\left|F\right|m_S\right)}
{\cosh^2\left(2\left|F\right|m_S\right) - \sin^2\frac{\theta}{2}}\right|_{\theta=\pi} -$$
$$-\left.\sinh\left(4\left|F\right|m_S\right)\frac{\frac{dm_S}{d\theta}
\left|F\right|\sinh\left(4\left|F\right|m_S\right) 
- \frac{\sin\theta}{4}}{\left(\cosh^2\left(2\left|F\right|m_S\right) 
- \sin^2\frac{\theta}{2}\right)^2}\right|_{\theta=\pi} =$$
$$= \left.\frac{dm_S}{d\theta}\right|_{\theta=\pi}4\left|F\right|
\cotanh\left(2\left|F\right|m_S\right)\left[1 - \cotanh\left(2\left|F
  \right|m_S\right)\right].
\label{AppX-XIX}
\end{equation}
Moving all the terms to the l.h.s. of the equation we find that either
\begin{equation}
1 - 4\left|F\right|\cotanh\left(2\left|F\right|m_S\right) \left[1 -
  \cotanh\left(2\left|F\right|m_S\right)\right] = 0,
\label{App-XX}
\end{equation}
or
\begin{equation}
\left.\frac{dm_S}{d\theta}\right|_{\theta=\pi} = 0.
\label{App-XXI}
\end{equation}
The first case is impossible, for the solution to the saddle-point
equation at $\theta=\pi$ imposes
\begin{equation}
m_S\left(\pi\right) = \cotanh\left(2\left|F\right|m_S\right),
\label{App-XIXI}
\end{equation}
which is non-zero and verifies
$$\left|m_S\right| \geq 1,$$ so the l.h.s. never vanishes, for the
second summand is always positive. Therefore, \eqref{App-XXI} applies
and the derivative vanishes at $\theta=\pi$.

%Stanley

%Michèle Le Bellac

%\bibitem{PhaseTQCD}
%H. Meyer-Ortmanns,
%\emph{Phase Transitions in Quantum Chromodynamics}
%[{\tt arXiv:hep-lat/9608098v1}]

\end{document}